\documentclass[letterpaper, 10 pt, conference]{ieeeconf}  % Comment this line out if you need a4paper

\IEEEoverridecommandlockouts                              % This command is only needed if
\usepackage{graphicx} % for pdf, bitmapped graphics files
\graphicspath{{data/}{images/}}
\usepackage{amsmath} % assumes amsmath package installed
\usepackage{amssymb}  % assumes amsmath package installed
\usepackage{multirow}
\usepackage{url}

\newcommand\invisiblesection[1]{%
\refstepcounter{section}%
\addcontentsline{toc}{section}{\protect\numberline{\thesection}#1}%
\sectionmark{#1}}

\title{\LARGE \bf
Reshaping the use of digital tools to fight malaria
}

\author{Sekou L. Remy, Oliver Bent, Nelson Bore% <-this % stops a space
\thanks{Sekou L. Remy and Nelson Bore are with IBM Research Africa, {\tt\small sekou@ke.ibm.com, nelsonbo@ke.ibm.com} \hspace{1cm}
Oliver Bent is a doctoral student at the University of Oxford and a visiting researcher at IBM Research Africa, {\tt\small oetbent@robots.ox.ac.uk} \hspace{.5cm}
\textbf{This work was presented at the Novartis Sponsored Symposium: \textit{Using digital tools to strengthen the malaria supply chain}, The Multilateral Initiative on Malaria, Dakar Senegal, April 2018.}
}%
}

\begin{document}

\maketitle
\thispagestyle{empty}
\pagestyle{empty}

%%%%%%%%%%%%%%%%%%%%%%%%%%%%%%%%%%%%%%%%%%%%%%%%%%%%%%%%%%%%%%%%%%%%%%%%%%%%%%%%
\invisiblesection{INTRODUCTION}
Disease modeling and simulation efforts for malaria control and eradication are among the most highly developed for any disease. 
Researchers fighting malaria have access to several transmission models developed by organizations across the globe \cite{smith2006mathematical,arifin2016spatial,GAMBHIR2017e638}.
This is an active research space, however it is still rare for researchers to access models and data from contributions that they did not perform themselves.
If highly trained researchers are not sharing and using models developed in the community, it becomes a further challenge for policy makers working in the control and eradication of malaria, who ultimately need to consume their scholarship.
The consensus is that at present it is not easy enough to marshal all simulation and modeling assets, as a team we present an approach to do so.

IBM Research Africa is one of IBM's solutions focused global labs.
With offices in Nairobi, Kenya and Johannesburg, South Africa, the lab exists on a continent where malaria is endemic.
Focused on Africa's grand challenges, our lab is interested in entering the fight to eradicate malaria by harnessing expertise in 1) advanced computing resources and 2) applied artificial intelligence (AI).
IBM is known for research and effective applications in these two domains, and we would like to leverage our resources to accelerate the computational assault on the disease.
There are two key areas that this expertise can be applied:
the first is improving the quality and nature of the models, and 
the second is improving the utilization of these models.

In the area of improving the models used, there is a need to generate better models, and to calibrate these models with the appropriate data.
The improvement in model complexity (for example capturing facets such as drug resistance) 
%and the existence of nuanced vaccination strategies
has the side-effect of making these models harder to execute, share, and for policy makers to consume.
The push to improve the quality of the data used to develop models also comes with challenges.
In our view these center on how to allocate resources to collect/generate this data, and how to effectively enable sharing of this data to inform users of the data (and the associated models).
%If these can be addressed, we envision that it would enable a better division of labor as  

Utilizing the generated data and models provides a distinct, yet complimentary challenge for the malaria research community.
Policy makers who will consume the output or even researchers who could externally validate them, must be able to execute/access the findings. 
The true value in the findings is also realized if the models are able to be executed at scale, and if the computers needed to acquire the results can be readily configured.
Finally, the required computational resources often are associated with significant operational overhead, and if they are not well designed or deployed, their use results in sub-optimal allocation of limited resources.

\section{PROBLEM}
Applying computational models and relevant data to interesting malaria settings is often difficult to configure (including acquisition of the requisite data), difficult to execute (especially at scientifically relevant scales), and difficult to interpret.

\section{APPROACH}
In \cite{nets} we demonstrated an approach which addresses this problem, with the emphasis on exploration and interpretation.
We developed an infrastructure by applying common computing abstractions in software development and deployment, and then applied three classes of algorithms to generate insight from the developed infrastructure.

\subsection{Infrastructure for Modeling and Computing}
\begin{figure}[ht]
\centering
\includegraphics[width=.95\columnwidth]{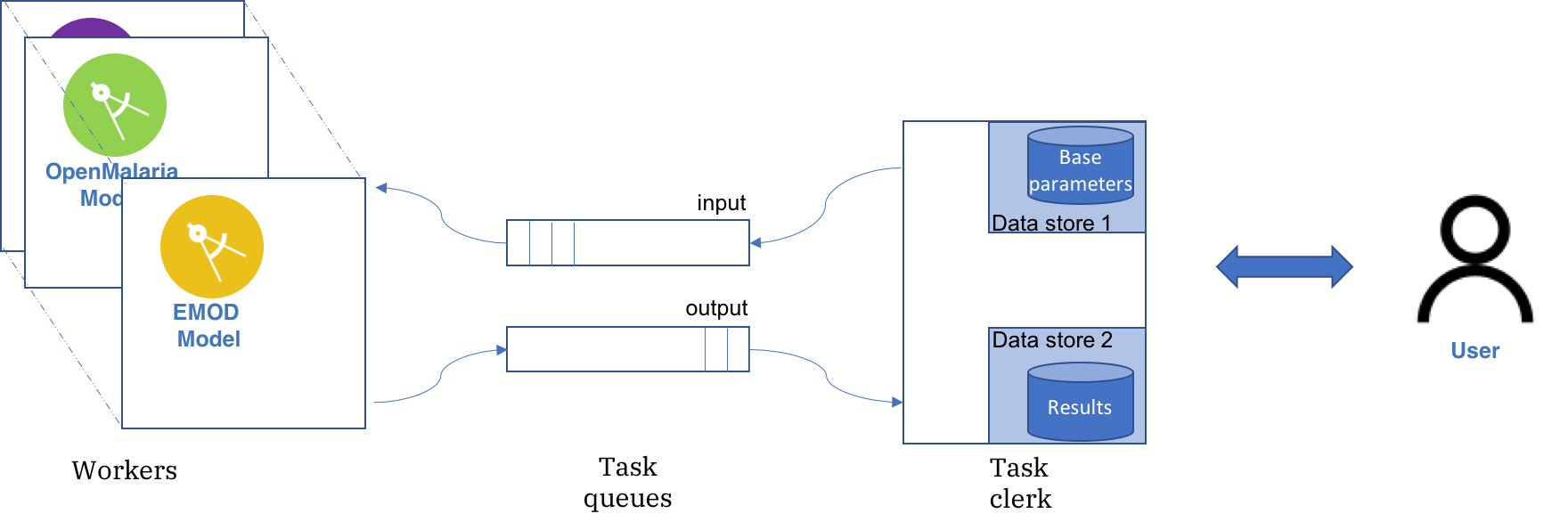}
\caption{Conceptual architecture of the infrastructure.}
\label{fig:arch}
\end{figure}

\subsubsection{Worker}
Using containers\footnote{https://hub.docker.com/r/kibnelson/openmalaria/}\footnote{https://hub.docker.com/r/slremy/emodworker/}, we have packaged malaria models in a manner which is easy to deploy at scale and in multiple types of computing environments.
In this container we also couple the model with software to communicate the desired input file, and to process the output files as needed.
Together, these tools are referred to as a \textit{worker}, and multiple workers can be deployed on a single machine or even distributed across the Internet.

\subsubsection{Task Clerk}
We permit users to define a specific instantiation of a model that they'd like to evaluate.
These parameters are use to ``germinate'' a seed input file for the model.
The resulting file is then sent as a task to the worker so that it can be processed.

At present the parameters define two properties of an intervention policy:
the portion of the households where insecticide treated bednets (ITN) are deployed, and
the portion of households where indoor residual spraying (IRS)is applied.
Other parameters provided by the model will be added in future work.

\subsubsection{Data store}
The seed input file, as well as the results for all evaluated policies are stored in a central repository.
The \textit{task clerk} and all the distributed \textit{workers} are connected via a common messaging fabric to this data store.
Results from the model's execution are converted to the cost per Disability Adjusted Life Year averted.
This measure is implemented to provide direct comparison with earlier published work \cite{Stuckey2014}
This data is also stored for subsequent reporting in the case that intervention is requested in the future.

\subsubsection{Task queue}
To tie each of the components together, we use a messaging fabric.
The current implementation harnesses AMQP as implemented in the RabbitMQ message queue.
The frontend posts jobs to a message queue which are subsequently picked up by idle \textit{workers} of the appropriate type.
When the \textit{worker} is complete, the results are posted to a different channel on the same queue.
This instantiation permits \textit{workers} to be deployed in a wide range of environments, with little requirements on coordination.

\subsection{Applying Artificial Intelligence}
The tools of Machine Learning or AI can now use the aforementioned infrastructure to assist with finding optimal or more optimal strategies as recommendations to malaria policy makers across the globe. AI can achieve this task in a way which has not been possible using all of the research capacity and compute we already have access to.
In \cite{nets} we framed the process of finding an optimal malaria policy as a stochastic multi-armed bandit problem, and implement three agent-based strategies to find optimal policies for a single environment.
The selected algorithms were already in existence, and the novelty of our early work resides in the application domain. However we believe the domain also provides an opportunity to justify the development of new algorithms which can better manage the complexity of the malaria problem.
\begin{figure}[ht]
\centering
\includegraphics[width=\columnwidth]{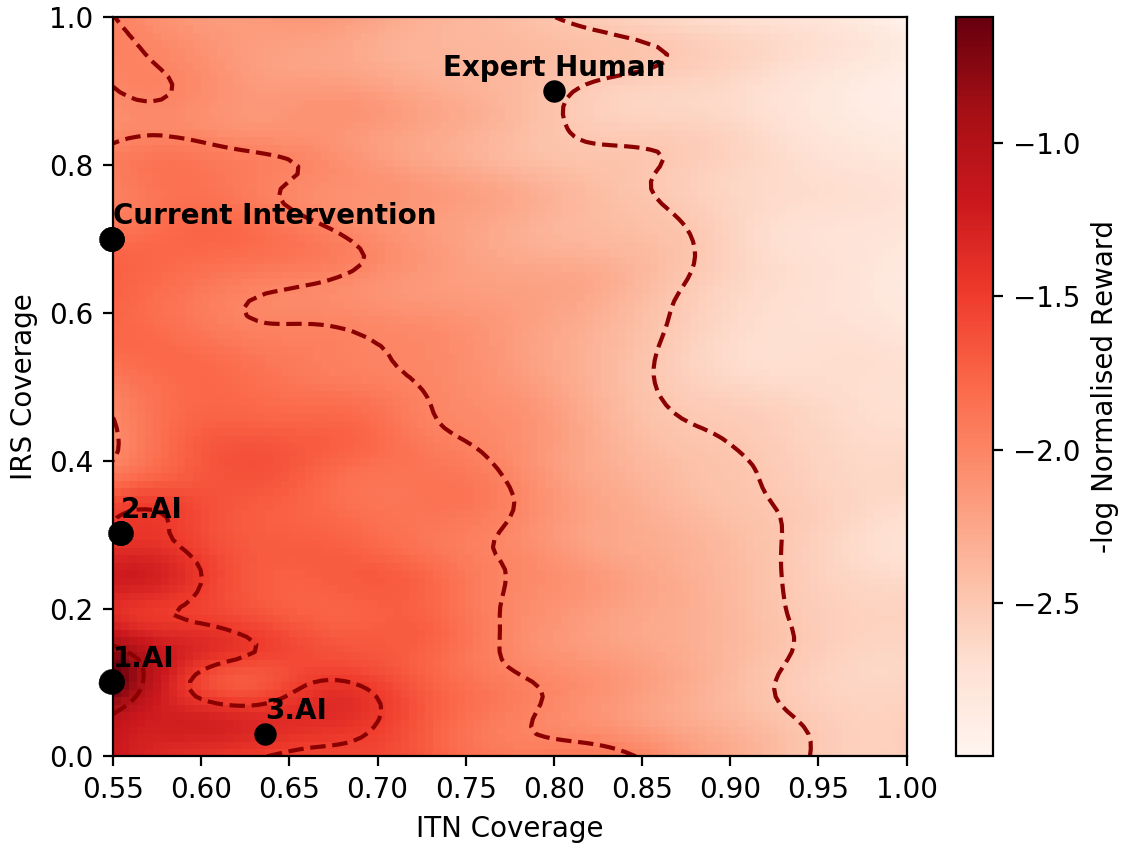}
\caption{Combined results from \cite{nets} showing AI (optimal), Current and Expert Human Solutions.}
\label{fig:entropy_external}
\end{figure}
Figure \ref{fig:entropy_external} depicts the results of using AI to question what is the most cost-efficient combination of two interventions (ITN and IRS).
The results are captured as surfaces where the optimal policies are dark red.
We do not assert that these results are better than the insight that is common in current policy making (although they do challenge the prevalent view).
Instead we consider these as the beginning of a dialog about both current policy and modeling approaches.
Our motivation is to use these techniques to explore more complex policy recommendations.

\section{SUMMARY}
In this manuscript we presented our approach to  marshal digital tools in the fight against malaria.
We describe our scalable infrastructure which leverages abstractions to support effective deployment of existing computational models and their associated data.
We also point to our earlier work leveraging such an infrastructure to effectively make more optimal policy recommendations with AI.

For the future of our work, we envision contributions in four complementary yet distinct areas:
\begin{enumerate}
\item Evidence-based policy improvement/implementation,
\item Augmented intelligence for complex decision making,
\item Better simulations, effectively using computation,
\item Transparent policy-based sharing of compute, data, models, and results towards global eradication.
\end{enumerate}
Moving forward, we are assembling a diverse, geographically distributed consortium of institutions to help shape what we hope to be a catalyst for the eradication of malaria.

%%%%%%%%%%%%%%%%%%%%%%%%%%%%%%%%%%%%%%%%%%%%%%%%%%%%%%%%%%%%%%%%%%%%%%%%%%%%%%%%

%%%%%%%%%%%%%%%%%%%%%%%%%%%%%%%%%%%%%%%%%%%%%%%%%%%%%%%%%%%%%%%%%%%%%%%%%%%%%%%%

%%%%%%%%%%%%%%%%%%%%%%%%%%%%%%%%%%%%%%%%%%%%%%%%%%%%%%%%%%%%%%%%%%%%%%%%%%%%%%%%
%\section*{APPENDIX}


\section*{ACKNOWLEDGMENT}
We are indebted to our managers Aisha Walcott-Bryant, Professor Stephen Roberts, and Komminist Weldemariam for their guidance and support.

%\section*{LEGAL}
%IBM,  the  IBM  logo,  and  ibm.com  are  trademarks  or  registered  trademarks  of  International  Business  Machines  Corp.,registered  in  many  jurisdictions  worldwide.
%Other  product and  service  names  might  be  trademarks  of  IBM  or  other companies.
%A  current  list  of  IBM  trademarks  is  available on  the  Web  at  ÓCopyright  and  trademark informationÓ  at \url{http://www.ibm.com/legal/copytrade.shtml}.

%%%%%%%%%%%%%%%%%%%%%%%%%%%%%%%%%%%%%%%%%%%%%%%%%%%%%%%%%%%%%%%%%%%%%%%%%%%%%%%%

%References are important to the reader; therefore, each citation must be complete and correct. If at all possible, references should be commonly available publications.

\bibliography{../../bibref/research_ref,root}
\bibliographystyle{IEEEtran}

\end{document}